\documentclass[prl,twocolumn,showpacs,preprintnumbers,amsmath,amssymb,superscriptaddress]{revtex4}
\usepackage{color}
\usepackage{graphicx}% Include figure files
\usepackage{dcolumn}% Align table columns on decimal point
\usepackage{bm}% bold math
\usepackage{ulem}

\begin{document}

\title{Precise Characterization of $^6$Li Feshbach Resonances \\
Using Trap-Sideband-Resolved RF Spectroscopy of Weakly Bound Molecules}

\author{G.\,Z\"urn}
\affiliation{Physikalisches Institut, Ruprecht-Karls-Universit\"at Heidelberg, 69120 Heidelberg, Germany}
\affiliation{Max-Planck-Institut f\" ur Kernphysik, Saupfercheckweg 1, 69117 Heidelberg, Germany}
 \author{T.\,Lompe}
 \email{lompe@physi.uni-heidelberg.de}
\affiliation{Physikalisches Institut, Ruprecht-Karls-Universit\"at Heidelberg, 69120 Heidelberg, Germany}
\affiliation{Max-Planck-Institut f\" ur Kernphysik, Saupfercheckweg 1, 69117 Heidelberg, Germany}
\affiliation{ExtreMe Matter Institute EMMI, GSI Helmholtzzentrum f\"ur Schwerionenforschung, 64291 Darmstadt, Germany}
 \author{A.\,N.\,Wenz}
\affiliation{Physikalisches Institut, Ruprecht-Karls-Universit\"at Heidelberg, 69120 Heidelberg, Germany}
\affiliation{Max-Planck-Institut f\" ur Kernphysik, Saupfercheckweg 1, 69117 Heidelberg, Germany}
\author{S.\,Jochim}
\affiliation{Physikalisches Institut, Ruprecht-Karls-Universit\"at Heidelberg, 69120 Heidelberg, Germany}
\affiliation{Max-Planck-Institut f\" ur Kernphysik, Saupfercheckweg 1, 69117 Heidelberg, Germany}
\affiliation{ExtreMe Matter Institute EMMI, GSI Helmholtzzentrum f\"ur Schwerionenforschung, 64291 Darmstadt, Germany}

\author{P.\,S.\,Julienne}
\affiliation{Joint Quantum Institute, NIST and the University of Maryland,
Gaithersburg, Maryland 20899-8423, USA}

\author{J.\,M.\,Hutson}
\email{J.M.Hutson@durham.ac.uk} \affiliation{Department of Chemistry, Joint Quantum Centre (JQC)
Durham/Newcastle, Durham University, South Road,
Durham, DH1~3LE, United Kingdom}



\begin{abstract}
We perform radio-frequency dissociation spectroscopy of weakly
bound $^6$Li$_2$ Feshbach molecules using low-density samples of about
30 molecules in an optical dipole trap. Combined with a high magnetic
field stability this allows us to resolve the discrete trap levels in
the RF dissociation spectra. This novel technique allows the binding
energy of Feshbach molecules to be determined with unprecedented
precision. We use these measurements as an input for a fit to the
$^6$Li scattering potential using coupled-channel calculations. From
this new potential, we determine the pole positions of the broad $^6$Li
Feshbach resonances with an accuracy better than $7 \times 10^{-4}$ of
the resonance widths. This eliminates the dominant uncertainty for
current precision measurements of the equation of state of strongly
interacting Fermi gases. As an important consequence, our results imply
a corrected value for the Bertsch parameter $\xi$ measured by Ku et
al.\ [Science 335, 563 (2012)], which is $\xi = 0.370(5)(8)$.
\end{abstract}


\maketitle

In the past few years, ultracold Fermi gases of neutral atoms have
become important benchmark systems for testing theories of strongly
interacting many-body systems \cite{Bloch2008}. This success is based
on two main factors. The first is that the physics of ultracold gases
is very well approximated by simple model Hamiltonians. These
Hamiltonians contain only a contact interaction, which can be described
by a single quantity, the scattering length $a$. The second is the
existence of Feshbach resonances in the interparticle scattering, which
cause the scattering length to diverge to $\pm \infty$ at certain
magnetic field values $B_0$ \cite{Chin2010}. This allows tuning of the
interparticle interactions by applying a homogeneous magnetic offset
field. Using such resonances, the properties of strongly interacting
Fermi gases have been investigated using a number of different
techniques, which range from radio-frequency (RF) spectroscopy
\cite{chin_pairing_gap, schunck_pair_size}, through studies of
collective oscillations \cite{innsbruck_oscillations,
john_thomas_oscillations}, to the detailed analysis of in-trap density
profiles \cite{salomon_eos, ueda_eos, zwierlein_eos}. However,
regardless  of which technique is used, all such measurements depend on
accurate knowledge of the properties of the Feshbach resonance that is
used to tune the interactions.

$^6$Li atoms in the three energetically lowest Zeeman sublevels of the
electronic ground state (labeled $\vert1\rangle$, $\vert2\rangle$ and
$\vert3\rangle$ following Ref.\ \cite{Bartenstein2005}) are widely used
to realize strongly interacting Fermi gases. The interactions
between atoms in the three different spin states are described by three
scattering lengths $a_{12}$, $a_{23}$ and $a_{13}$, which can all be
tuned using broad Feshbach resonances located at magnetic fields of
about 800\,G with resonance widths of up to 300\,G \footnote{Units of
gauss rather than Tesla, the accepted SI unit of magnetic field, are
used in this paper to conform to the conventional usage of this
field.}. These resonances have been used to create the best known
realization of a Fermi gas with diverging scattering length, which is a
valuable benchmark system for many-body theories. How well this
benchmark system can be realized is currently limited by the accuracy
of the previous determination of the resonance positions, which was
$\lesssim 1.5$\,G \cite{Bartenstein2005}.
% and therefore already better than $1\,\%$ of the resonance widths.
%The positions of these resonances were determined by Bartenstein et al.\
%with an accuracy of about 1\,G \cite{Bartenstein2005}, which corresponds
%to less than $1\,\%$ of the resonance widths.
Recent studies of the equation of state (EoS) of strongly interacting
Fermi gases have reached a level of precision at which they are limited
by these uncertainties in the resonance positions. An important example
is measurements recently performed by Nascimb{\'e}ne et al.\
\cite{salomon_eos} and Ku et al.\ \cite{zwierlein_eos} with the goal of
measuring the EoS at the point where the scattering length diverges to
$\pm \infty$. In this so-called unitary limit the scattering length
drops out of the problem, leaving the interparticle spacing as the only
remaining length scale.  At zero temperature this has the consequence
that all extensive quantities of the unitary Fermi gas are given by
their values for a noninteracting system rescaled by a universal
numerical constant $\xi$, known as the Bertsch parameter
\cite{Baker2001}. Ku et al.\ determined this parameter to be $\xi =
0.376 \pm 0.004$, providing a precision measurement that can serve as a
test for theories in such different fields as cold gases, nuclear
physics and the physics of neutron stars. However, if the measurement
is performed at a finite value of the scattering length, it leads to
systematic errors. The error in $\xi$ resulting from the $1.5\,$G
uncertainty in the resonance position determined by Bartenstein et al.\
is about  $2\,\%$ and is the largest error contribution
\cite{zwierlein_eos}. This clearly illustrates the necessity of a new,
more accurate determination of the properties of the $^6$Li Feshbach
resonances.

\begin{figure} [tb]
\centering
	\includegraphics [width= 8cm] {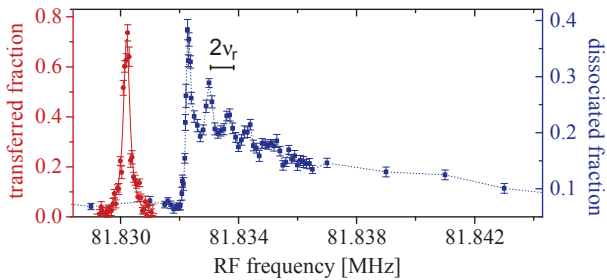}
	\caption{RF spectra for the free-free (red dots, left axis) and bound-free
(blue squares, right axis) spectra at a magnetic field of $B=811.139$\,G.
The line shape of the free-free transition is well described by a Lorentzian
(solid line). The bound-free spectrum
%is given by the wave function overlap between the initial molecular
%state and the possible final states of the dissociated atoms in the
%trapping potential. This allows us to observe
shows distinct peaks spaced by $2 \nu_{\rm r}$, corresponding to
different radial trap levels. The errors are the standard errors of the mean
of about $40$
individual measurements. }
	\label{fig:sample_spectrum}
\end{figure}	

In this work we determine the positions of the broad $^6$Li Feshbach
resonances with an accuracy of 80\,mG, which corresponds to less than
$7 \times 10^{-4}$ of the resonance widths. To achieve this we make use
of the fact that every Feshbach resonance is related to a weakly bound
dimer state. Close to the resonance the binding energy of the dimer is
approximately related to the scattering length by the universal
relation $E_{\rm b} = \hbar^2/m a^2$, where $m$ is the mass of one atom
\cite{Chin2010}. Thus we can obtain information about the $^6$Li
Feshbach resonances by measuring the binding energy of such a weakly
bound dimer state for different values of the magnetic field. However,
the universal relationship is not accurate enough for quantitative
interpretation, and in the present work we fit the measured binding
energies to determine a new model interaction potential for $^6$Li
using coupled-channel calculations. This new potential in turn provides
$a(B)$ as a function of magnetic field $B$ and allows us to
characterize the Feshbach resonances to high precision.

The most precise method currently available to measure the binding
energy of these dimers is RF spectroscopy \cite{jin_molecules,
Bartenstein2005}. This technique is based on applying an
RF pulse to a
gas of atoms to drive them from an initial hyperfine state $\vert
i\rangle$ to a final state $\vert f\rangle$. For a sample of molecules
one can either drive a transition to another weakly bound dimer state
(bound-bound transition) or dissociate the dimer into two free atoms
(bound-free transition). In either case the transition frequency is
shifted from the free-free transition by the difference in the binding
energies of the initial and final states. However, the transition
frequency is also affected by the difference in the mean-field energies
of the initial and final states. To avoid this systematic error,
measurements of the dimer binding energy must be performed in a regime
where the scattering length is much smaller than the interparticle
spacing, i.e.\ $na^3 \ll 1$. In previous experiments this could be
achieved only for relatively small values of $a \lesssim 2000\,a_0$, as
the experimentally achievable densities were limited to $n \gtrsim
10^{13}$\,molecules/cm$^3$. Accordingly, the smallest binding energies
that could be measured were on the order of 
$E_{\rm b} \simeq h \times 100$\,kHz, which resulted in a large uncertainty in the fitted
resonance position.

\begin{figure} [b]
\centering
	\includegraphics [width= 7cm] {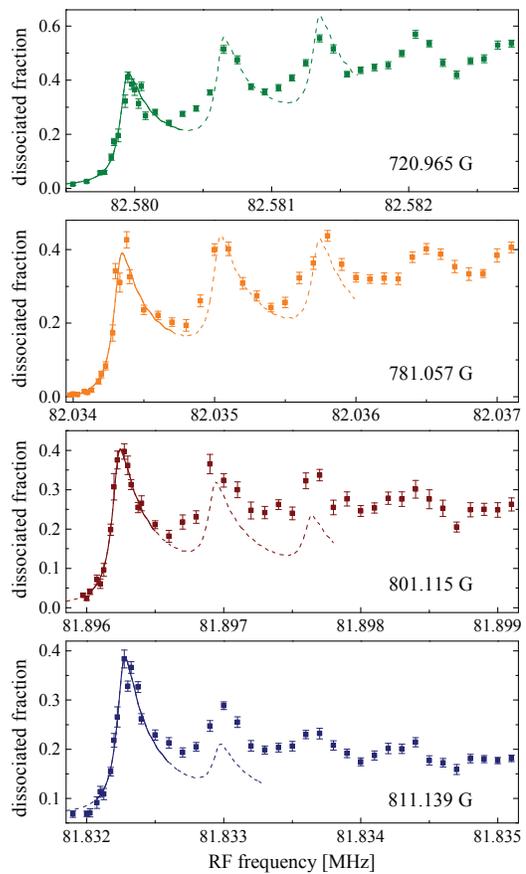}
	\caption{Molecular dissociation spectra at four different magnetic fields.
The lines show fits according to the model described in the text, with the
solid parts indicating the range of the data points included in the fit.
 }
\label{fig:all_spectra}
\end{figure}	

\begin{table*} [tbh]
\begin{center}
\small
  \begin{tabular}{c | c | c	| r | c |r} \hline\hline
 Magnetic field & Free-free transition & Bound-free transition &  Dissociation frequency & Confinement shift      & Binding energy/$h$ \\
     $B$ (stat.)  & $\nu_{\rm ff}$ (stat.) & $\nu_{\rm bf}$ (stat.)(sys.)   & $\nu_{\rm bf}-\nu_{\rm ff}$ (stat.)(sys.) &
     %$\delta\nu$ (stat.)(sys.) &
     $\nu_{\rm cs}=\nu_0 (\rm{sys.}) + \nu_i (\rm{sys.})$  & $\nu_{E_{\rm b}}$ (stat.)(sys.) \\
     (G)  & (MHz)  & (MHz)	&   \multicolumn{1}{c|}{(kHz)}  & (kHz) & \multicolumn{1}{c}{(kHz)} \\ \hline
    811.139 (1) &81.\,830 115  \, (3) &81.\,832 271 \, (7)(8)	& 2.156 \, (8)(16)   &$0.367(3) - 0.014(1)$  & 1.803  \,  (8)(17)  \\
	801.115 (5) &81.\,891 539 (33)	  &81.\,896 236 \, (3)(8)	&4.697 (33)(16)   	 &$0.367(3) - 0.011(1)$  & 4.341     (33)(17)  \\
    781.057 (1) &82.\,019 823  \, (1) &82.\,034 336 \, (6)(8)	&14.513 \, (6)(16)   &$0.367(3) - 0.011(1)$  & 14.157 \,  (7)(17)  \\
	720.965 (1) &82.\,452 482  \, (2) &82.\,579 943  (13)(8)	&127.461  (13)(16)   &$0.367(3) - 0.021(1)$  & 127.115	 (14)(17)  \\
  \hline\hline
  \end{tabular}
\caption{Measured frequencies and resulting binding energies at different magnetic fields.
The dissociation frequency $\delta\nu$ is obtained by subtracting the
free-free transition frequency $\nu_{\rm ff}$ from the bound-free
transition frequency $\nu_{\rm bf}$.
To obtain the binding energy we subtract the confinement induced
frequency shift $\nu_{\rm cs}$ from the dissociation frequency
$\delta\nu$.
The different contributions to the confinement induced shift $\nu_{\rm cs}$ and the
statistical
(stat.)
and systematic
(sys.)
errors are discussed in the Supplemental
Material \cite{SOM}.}
%This shift has two contributions: The first is the shift $\nu_0 = \nu_{\rm r}+ \frac{1}{2} \nu_{\rm ax}$ caused by the zero-point energy of the relative motion of the dissociated particles. The second contribution is the interaction shift $\nu_i$ caused by the confinement of the molecules and the interactions between the dissociated particles \cite{idziaszek2006}. A complete discussion of the statistical and systematic errors can be found in the Supplemental Material \cite{SOM}.}
  \label{table_RF_results}
\end{center}
\end {table*}

We use the techniques we have developed to prepare and detect
few-particle systems \cite{Serwane2011} to create very dilute samples
of molecules. This allows us to perform RF spectroscopy of dimers with
much smaller binding energies, and thus measure much closer to the
resonance. We start from a small Bose-Einstein condensate of about
$10^3$ $|12\rangle$ molecules, trapped in a small-volume optical dipole
trap at a magnetic field of $760$\,G. Subsequently we reduce the
particle number to about 30 molecules by applying the spilling
technique developed in Ref.\ \cite{Serwane2011}. We then superimpose a
large-volume optical dipole trap with trap frequencies of $\nu_{\rm r}
= \omega_{\rm r}/ 2\pi= 349(3)$\,Hz and $\nu_{\rm ax} = \omega_{\rm
ax}/2\pi = 35(1)$\,Hz in the radial and axial directions, respectively.
To transfer the molecules into this shallow dipole trap we suddenly
switch off the microtrap. This nonadiabatic release results in a mean
kinetic energy per particle of $\geq 0.4\,\mu$K and therefore a final
molecular peak density of $n \leq 10^9$\,molecules/cm$^3$ which greatly
reduces density-dependent shifts of the RF transition \cite{SOM}.
%This corresponds to an interparticle spacing of about 10\,$\mu$m.

To measure the bound-free spectra we first perform a $10\,$ms
ramp from the magnetic field of $760$\,G at which we prepare the
sample to the magnetic field of interest and wait for another $5$\,ms. This
time is long enough for the magnetic offset field to stabilize
to an uncertainty of typically $1 \,$mG, %\times 10^{-3}\,$G,
but short enough to avoid collisional dissociation of a significant
fraction of molecules. We then apply a rectangular RF pulse of $10$\,ms
duration to dissociate a fraction of the $\vert12\rangle$ molecules
into free atoms in states $\vert1\rangle$ and $\vert3\rangle$. By
measuring the number of these unbound atoms as a function of the RF
frequency, we obtain spectra as shown by blue dots in Fig.
\ref{fig:sample_spectrum}. To limit saturation effects, we choose the
pulse power such that we dissociate at most $30\,\%$ to $40\,\%$ of the
molecules.
%From these spectra we extract the frequency $\nu_{\rm bf}$
%of the bound-free transition.

To measure the frequency of the free-free transition we prepare a
spin-polarized Fermi gas of atoms in state $\vert2\rangle$ and drive the RF
transition to state $\vert 3\rangle$ (red dots in Fig.
\ref{fig:sample_spectrum}). We do this before and after the molecule
dissociation measurement and use the weighted mean $\nu_{\rm ff}$ of
the two measurements \cite{SOM}. From this we can
also determine the magnetic field using the Breit-Rabi formula.

The profile of the bound-free spectrum is determined by the overlap between the
wave functions of the initial molecular state $\psi_{i}$ and of the accessible
final states $\psi_{f}$ \cite{chin_julienne_lineshape}. As the RF photons carry
only negligible momentum, there is no net momentum transfer to the system and
therefore the RF pulse can affect only the relative motion of the two atoms.
The transition rate between the initial and final state is thus proportional to
$\left\vert \langle \psi_{i} (r) \vert \psi_{f}(r) \rangle \right\vert^2$,
where $r$ is the distance between the two atoms. For a continuum of final
states  the resulting asymmetric line shape  is given by the momentum
distribution of the initial molecular state \cite{chin_julienne_lineshape}. In
a confining potential, however, the final states are the discrete energy levels
of the trap and the profile is determined by the overlap between the molecular
state and the trap states. If the experimental resolution is insufficient to
resolve the trap levels, as was the case in previous experiments
\cite{jin_molecules, Bartenstein2005, chin_pairing_gap, schunck_pair_size}, the
final state can be well described by a continuum. In our case the resolution is
high enough to resolve the radial trap levels (see Fig.
\ref{fig:all_spectra}).

For the initial molecular state, the long-range part of the wave function
 for the relative motion is well described by $\psi_{i}(r) \simeq
e^{- r/a}$, where $a$ is the scattering length. The final states are
the levels of our optical trap, which we approximate as harmonic
oscillator levels. Since the initial state  is symmetric, only the
symmetric harmonic oscillator levels ($n_{\rm ho}$=0,2,4,...)
contribute. Calculating the wave function overlap results in a spectrum
of delta functions of different heights located at $\nu_{\rm bf} + p
\nu_{\rm r}+ q \nu_{\rm ax}$, where $\nu_{\rm bf}$ is the frequency of
the bound-free transition and $p,q$ are non-negative even integers. To
fit our measured spectra, we convolute this spectral function with the
line shape of the free-free transition, which we approximate by a
Lorentzian with a FWHM of $122$\,Hz. Because of this finite resolution only
the radial peaks are resolved. The free parameters of the fits are
$\nu_{\rm bf}$, the overall amplitude, and a small offset in the atom
number arising from collisional dissociation of molecules
\cite{chin2004chemicalequilibrium}. To determine $\nu_{\rm bf}$ we fit
the lowest radial peak at each field (solid lines in
Fig.~\ref{fig:all_spectra}) \footnote{ As we chose to dissociate a
relatively large fraction of the molecules ($\simeq 40\,\%$ for the
first peak) to obtain a good signal-to-noise ratio for this fit, the
absolute height of the peaks is affected by saturation. If the
different peaks have similar heights, as is the case at $B=781\,$G,
saturation affects them equally and the model shows quantitative
agreement for all peaks. If the peak heights differ strongly the model
shows only qualitative agreement for the peaks corresponding to final
states with higher energy (dashed lines in fig.~\ref{fig:all_spectra}).
}. The molecular binding energies, obtained by subtracting the
confinement-induced frequency shifts \cite{idziaszek2006, SOM} from the
dissociation frequencies $\nu_{\rm bf}-\nu_{\rm ff}$, are given in
table \ref{table_RF_results}.

To fit the experimental results and extract the position of the broad resonance
pole, we use a coupled-channel model similar to that of Refs.\
\cite{O'Hara:Li-zero:2002} and \cite{Bartenstein2005}. The interaction
potentials are constructed using the short-range singlet potential of Ref.\
\cite{Cote:1994} and the short-range triplet potential of Ref.\
\cite{Linton:1999}, joined at long range onto potentials based on the
dispersion coefficients of Ref.\ \cite{Yan:1996} and the exchange function of
Ref.\ \cite{Cote:1994}. The interatomic spin-dipolar interaction is taken to
follow its long-range ($r^{-3}$) form at all distances. The singlet and triplet
scattering lengths are adjusted by making small changes to the repulsive walls
of the singlet and triplet potentials with parameters $S_0$ and $S_1$.
Scattering calculations are carried out using the MOLSCAT package
\cite{molscat:2011}, and bound-state calculations using the companion package
BOUND \cite{Hutson:bound:2011, Hutson:Cs2:2008}. MOLSCAT can converge directly
on the positions of poles and zeroes in the scattering length. We carried out
least-squares fits to the new binding energy measurements described above,
together with the two bound-bound spectroscopic frequencies of Ref.\
\cite{Bartenstein2005} at 661.436~G and 676.090~G, the magnetic field near
527~G where the scattering length passes through zero \cite{Du2008}, and
the position of the narrow resonance near 543~G
\cite{hazlett2012}. The least-squares fits were
carried out using the interactive package I-NoLLS \cite{I-NoLLS}.

A two-parameter fit using only $S_0$ and $S_1$ proved capable of giving
a good fit to all the experimental results {\em except} the position of
the narrow resonance. This fit placed the narrow resonance about 0.12~G
to high field of its experimental position. This discrepancy could be
resolved by introducing a third parameter in a variety of ways, such as
scaling the exchange potential or changing the value of the exponent
parameter $\beta$ in the exchange potential. However, in the absence of
a good theoretical justification for the extra parameters, and since
introducing them had little effect on the parameters of the resonances
near 800~G, we ultimately chose a two-parameter fit, excluding the data
point for the pole of the narrow resonance, as the most reliable for
our purpose. To estimate the uncertainties in the pole positions and
derived parameters, we repeated the fits using binding energies at the
upper and lower limits of the systematic uncertainties and used the
range of predictions from the various fits to estimate the
model dependence.

The quality of fit and the key quantities calculated from the best-fit
(two-parameter) potential are summarized in Table \ref{res_params}.
Tabulations of $a(B)$ for the best-fit potential are given in the
Supplemental Material \cite{SOM}.

\begin{table}
 \setlength{\tabcolsep}{0mm}
\begin{tabular}{r  l | r l | r  l | r  l} \hline\hline
\multicolumn{2}{c|}{}& \multicolumn{2}{c|}{Fit Ref.\ \cite{Bartenstein2005}}  & \multicolumn{2}{c|}{Present fit} & \multicolumn{2}{c}{Experiment} \\ \hline
\multicolumn{2}{c|}{$\nu_{\rm b,12}-\nu_{\rm b,13}+\nu_{\rm ff}$} &  \multicolumn{2}{c|}{\,83 664.0(10)\,} & \multicolumn{2}{c|}{83 665.9(3)}
     &  \multicolumn{2}{c}{\,83 664.5(10) } \\
\multicolumn{2}{c|}{at 661.436 G }& \multicolumn{2}{c|}{} & \multicolumn{2}{c|}{} & \multicolumn{2}{r}{\cite{Bartenstein2005}}   \\
\multicolumn{2}{c|}{$\nu_{\rm b,12}-\nu_{\rm b,13}+\nu_{\rm ff}$} & \multicolumn{2}{c|}{\,83 297.3(10)\,} & \multicolumn{2}{c|}{83 297.3(3)} & \multicolumn{2}{c}{\,83 296.6(10)}  \\
\multicolumn{2}{c|}{at 676.090 G} & \multicolumn{2}{c|}{} & \multicolumn{2}{c|}{} & \multicolumn{2}{r}{\cite{Bartenstein2005}}  \\
\multicolumn{2}{c|}{$\nu_{\rm b,12}$ at 720.965 G } &\multicolumn{2}{c|}{}  &\,127&.115(17)\,						&\;\,127&.115(31)       \\
\multicolumn{2}{c|}{$\nu_{\rm b,12}$ at 781.057 G } &\multicolumn{2}{c|}{}  &   14&.103(26)  						&   14&.157(24)         \\
\multicolumn{2}{c|}{$\nu_{\rm b,12}$ at 801.115 G } &\multicolumn{2}{c|}{}  &    4&.342(17)  						&    4&.341(50)         \\
\multicolumn{2}{c|}{$\nu_{\rm b,12}$ at 811.139 G } &\multicolumn{2}{c|}{}  &    1&.828(11)  						&    1&.803(25)         \\
\multicolumn{2}{c|}{Zero in $a_{12}$ }              &\multicolumn{2}{c|}{}  &   527&.32(25)  						&  527&.5(2)\cite{Du2008}  			\\
\multicolumn{2}{c|}{Narrow pole in $a_{12}$\, }     &\multicolumn{2}{c|}{}  &   543&.41(12)  						&  543&.286(3)			\\
\multicolumn{2}{c|}{$a_{\rm s}$  }                        & \multicolumn{2}{c|}{45.167(8)}& 45&.154(10)       &\multicolumn{2}{r}{\cite{hazlett2012} }   \\
\multicolumn{2}{c|}{$a_{\rm t}$  }                        &\;$-$2140&(18) & \;$-$2113&(2) &\multicolumn{2}{c}{} \\
\hline
\end{tabular}
\begin{tabular}{l | c c | c  c | c  c}
\hline
   & \multicolumn{2}{c|}{ pole (G) } &\multicolumn{2}{c|}{ $\Delta$ (G) } &\multicolumn{2}{c}{ $a_{\rm bg}$ ($a_0$) } \\
   &\! Ref.\cite{Bartenstein2005} &\! Present\,fit \! &\! Ref.\cite{Bartenstein2005} &\! Present\,fit \!  &\! Ref.\cite{Bartenstein2005} &\! Present\,fit \\
\hline
$|12\rangle$\,  & \, 834.15 & \, 832.18(8)  & \, 300 \! \ & \, $-262.3$(3)  & \, $-1405$  & \, $-1582$(1)   \\
$|13\rangle$\,  & \, 690.43 & \, 689.68(8)  & \, 122.3    & \, $-166.6$(3)  & \, $-1727$  & \, $-1770$(5)   \\
$|23\rangle$\,  & \, 811.22 & \, 809.76(5)  & \, 222.3    & \, $-200.2$(5)  & \, $-1490$  & \, $-1642$(5)   \\
\hline \hline
\end{tabular}

%& & pole & &\Delta  & & \\
%\;\; \; \; \; \, \,&$|12\rangle$ pole &  834&.149   &    832&.18(8)      			   &\multicolumn{2}{c}{}  \\
%&$|12\rangle$ $\Delta$       &  300&        &    $-$262&.3(3)      								 &\multicolumn{2}{c}{}  \\
%&$|12\rangle$ $a_{\rm bg}$   & $-$1405&     &  $-$1582&(1)										   &\multicolumn{2}{c}{}  \\
%&$|13\rangle$ pole           &   690&.43    &    689&.68(8)      								 &\multicolumn{2}{c}{}  \\
%&$|13\rangle$ $\Delta$       &   122&.3     &    $-$116&.6(3)       								 &\multicolumn{2}{c}{}  \\
%&$|13\rangle$ $a_{\rm bg}$   & $-$1727&     &  $-$1770&(5) 											 &\multicolumn{2}{c}{}  \\
%&$|23\rangle$ pole           &   811&.22    &    809&.76(5)      								 &\multicolumn{2}{c}{}  \\
%&$|23\rangle$ $\Delta$       &   222&.3     &  $-$200&.2(5)       								 &\multicolumn{2}{c}{}  \\
%&$|23\rangle$ $a_{\rm bg}$   & $-$1490&     &  $-$1642&(5) 											 &\multicolumn{2}{c}{}  \\ \hline\hline
%\end{tabular}
\caption{Quality of fit between coupled-channel calculations on the
best-fit two-parameter $^6$Li potential and the experiments, together
with key derived quantities calculated using the potential. The
quantities in parentheses are estimates of the model dependence,
including the effect of the systematic errors in the binding energies
in Table \ref{table_RF_results}. All frequencies are given in kHz, all
lengths in bohr and all magnetic fields in G. The $\Delta$ and $a_{\rm
bg}$ values are obtained from {\em local} fits to $a(B)$ near the
resonance and do not correctly reproduce the positions of the zeroes in
$a(B)$.} \label{res_params}
\end{table}

With these results, the uncertainty in the positions of the broad
$^6$Li Feshbach resonances is no longer a limiting factor for current
experiments. Using our new calibration of $a(B)$ it is possible to
address systematic errors in recent experiments which were caused by
the inaccuracy of the previous determination of the resonance
positions. The most striking example of this is the determination of
the Bertsch parameter $\xi$ by Ku et al.\ \cite{zwierlein_eos}, which
was performed using a mixture of $^6$Li atoms in states $\vert
1\rangle$ and $\vert2\rangle$ at a magnetic field of 834.15\,G. At this
field, our best-fit potential gives $a(B)=-2.124(80)\times10^5\,a_0$
and effective range $r_{\rm eff}=87.03(1)$\,$a_0$. The difference
between the EoS at unitarity and the EoS measured at this finite value
of the scattering length may be obtained by using Tan's contact $C(a)$
\cite{Tan2008, zwierlein_eos}. This gives a corrected value for the
normalized zero-temperature chemical potential $\mu / E_F$ at
unitarity, which in turn gives a revised value of the Bertsch parameter
$\xi = 0.370 (5) (8)$ \cite{zwierlein_xi}. Here the first parenthesis
denotes the statistical error, while the second gives the systematic
uncertainty of the corrected value \footnote{This systematic
uncertainty is estimated from the difference between the corrected
values for the chemical potential $\mu /E_F = 0.370$, energy
$E/E_F=0.362$ and free energy $F/E_F=0.375$ of the unitary Fermi gas,
which should all converge to $\xi$ for $T \rightarrow 0$.}.

In this work we have established a new technique to measure the binding
energy of weakly bound molecules by performing trap-sideband-resolved
RF spectroscopy. By creating very dilute samples of molecules we have
greatly reduced density-dependent shifts of the RF transitions, which
has allowed us to perform spectroscopy of extremely weakly bound
molecules.  Using these techniques we have measured the binding energy
of $^6$Li Feshbach molecules with binding energies as low as $h \times
2$\,kHz with an accuracy better than $h \times 50$\,Hz, which is a
40-fold improvement compared to previous measurements
\cite{Bartenstein2005}. From these binding energies we have determined
the positions of the broad $^6$Li Feshbach resonances with an accuracy
of $80$\,mG using a coupled-channels calculation. This removes one of
the major limiting factors for precision studies of strongly
interacting Fermi gases. 

The authors thank M. Ku and M. Zwierlein for providing the corrected
value of $\xi$ as well as enlightening discussions. The authors
gratefully acknowledge support from IMPRS-QD, Helmholtz Alliance
HA216/EMMI, the Heidelberg Center for Quantum Dynamics, ERC Starting
Grant No. 279697, EPSRC, AFSOR MURI Grant No. FA9550-09-1-0617, and EOARD Grant No. FA8655-10-1-3033.

%\bibliography{paper}

\cleardoublepage

\section{Supplemental material}
\subsection{Determination of the dissociation frequency }%\cite{Zuerntakenthesis2012}}

To determine the dissociation frequency we measure the bound-free transition frequency $\nu_{\text{bf}}$  and the free-free transition frequency $\nu_{\text{ff}}$. To check for shifts of the RF-transitions during the experiment we measure the free-free transition before ($\nu_{\text{ff1}}$) and after ($\nu_{\text{ff2}}$) the molecule dissociation measurement and use the weighted mean $\nu_{\text{ff}}$ of both measurements. The dissociation frequency is then given by $\delta \nu = \nu_{\text{bf}} - \nu_{\text{ff}}$. The magnetic field is calibrated by inserting $\nu_{\text{ff}}$ into the Breit-Rabi-Formula \cite{BreitRabi1931}. 
To determine the density dependent shift $\Delta_{\nu_{\text{density}}}$ we increase the particle number from 30 to about 200 molecules. We find a shift of $\delta \nu$ which is smaller than 50\,Hz, from which we estimate a shift of less than $0.125\,$Hz per particle in a linear approximation. Therefore the systematic uncertainty due to density effects is $\Delta_{\nu_{\text{density}}} = 8\,$\,Hz for a sample of $30$ molecules ($60$ atoms). All parameters involved in the determination of the dissociation frequencies are listed in TABLE \ref{table free-free and bound-free}.

\vspace{1cm}
\begin {table*}[th!] 
  \begin{tabular}{ c | c |c|c |c| r| }
  magn. field & free-free 1st      &  free-free 2nd      & free-free weighted mean & bound-free transition& dissociation freq.\\
		B [G]        & $\nu_{\text{ff1}}$ [MHz]  & $\nu_{\text{ff2}}$ [MHz]   & $\nu_{\text{ff}}$ [MHz]   & $\nu_{\text{bf}}$ [MHz]         & $\delta\nu$ [kHz]\\ \hline
    811.139 (1) &81.\,830 120 \ (8)  &81.\,830 113  \ (5)  &81.\,830 115  \ (3) &  81.\,832 271 \ (7)(8)& 2.156 \ (8)(16)\\ 
		801.115 (5) &81.\,891 515 \ (3)  &81.\,891 583  \ (4)  &81.\,891 539 (33)		&  81.\,896 236 \ (3)(8)& 4.697 (33)(16) \\  
    781.057 (1) &82.\,019 822 \ (2)  &82.\,019 824  \ (3)  &82.\,019 823  \ (1)	&  82.\,034 336 \ (6)(8)& 14.513 \ (6)(16) \\  
		720.965 (1) &82.\,452 484 \ (4)  &82.\,452 479  \ (5)	 &82.\,452 482  \ (2)	&  82.\,579 943 (13)(8)& 127.461  (13)(16) \\ 
  \end {tabular}
	  \caption{\textbf{Transition and dissociation frequencies \cite{Zuerntakenthesis2012}.} The errors $\sigma_{\nu_{\text{ff1}}}$ and $\sigma_{\nu_{\text{ff2}}}$ of the measured free-free transition frequencies are the statistical errors of the fit to the free-free transition. $\nu_{\text{ff}}=\frac{\sum_{i} \frac{1}{\sigma_{\nu_{\text{ff}i}}^2}\nu_{\text{ff}i}}{\sum_{i} \frac{1}{\sigma_{\nu_{\text{ff}i}}^2} }$ is the weighted mean of $\nu_{\text{ff1}}$ and $\nu_{\text{ff2}}$ with error
		$\sigma_{\nu_{\text{ff}}}=\sqrt{\frac{\sum_{i} \frac{1}{\sigma_{\nu_{\text{ff}i}}^2}( \nu_{\text{ff}i} - \nu_{\text{ff}})^2 } {\sum_{i} \frac{1}{\sigma_{\nu_{\text{ff}i}}^2} }}$.
The magnetic field error $\sigma_{B}$ is the error resulting from the statistical error $\sigma_{\nu_{\text{ff}}}$. 
For the bound-free transition frequency $\nu_{\text{bf}}$ the first parenthesis gives the statistical error $\sigma_{\nu_{\text{bf}}}$ of a Lorentzian fit to the 
rising slope of the first peak of the spectrum. The second parenthesis gives the systematic error $\Delta_{\nu_{\text{model}}}$ of the fit, which we estimate by the difference between the fitted frequency using either a Lorentzian lineshape or a Gaussian lineshape to describe the transition peak into a single trap sideband.
For the dissociation frequency $\delta \nu$ the statistical error (first parenthesis) is obtained by quadratic addition of $\sigma_{\nu_{\text{ff}}}$ and $\sigma_{\nu_{\text{bf}}}$, while the systematic error is the sum of $\Delta_{\nu_{\text{model}}}$ and the systematic uncertainty due to density dependent shifts $\Delta_{\nu_{\text{density}}}$ (see text).
% (Breit-Rabi formula); 1st of $\text{\text{err}}_{\nu_{\text{bf}}}$: statistical error $\sigma_{\nu_{\text{bf}}}$: fit error of single Lorentzian fit to first slope of dissociation spectrum,  2nd of $\text{err}_{\nu_{\text{bf}}}$ : systematic error $\Delta_{\nu_{\text{model}}}$: max. of model dependent shift (Lorentz vs. Gauss); 1st of $\text{\text{err}}_{\delta\nu}$ : statistical error $\sigma_{\delta\nu}$: quadratic addition of $\sigma_{\nu_{\text{ff}}}$ and $\sigma_{\nu_{\text{bf}}}$,  2nd of $\text{\text{err}}_{\delta\nu}$: $\Delta_{\delta\nu}$ systematic error: $\Delta_{\nu_{\text{model}}}$+$\Delta_{\nu_{\text{density}}}$.  Taken from \cite{Zuern2012} and adapted.
		}
  \label{table free-free and bound-free}
\end {table*} 

\begin {table*}[th!]
  \begin{tabular}{ r | r |c| c|c|c|r|}
  dissociation freq.  &initial $a$  & final $a$& initial cs shift& final cs shift& cs shift      & binding energy/h \\
		$\delta\nu$ [kHz] & $a_{12}$  [$10^3\,$bohr] & $a_{13}$  [$10^3\,$bohr] & $\nu_{\text{cs-i}}$ [kHz]& $\nu_{\text{cs-f}}$ [kHz] & $\nu_{\text{cs}}$ [kHz]  & $\nu_{E_b}$ [kHz] \\ \hline
    2.156 \ (8)(16)   & 18.34   & -3.54(1) \ & 0.006& 0.359(1) & 0.353 (3)(1)  & 1.803 \, (8)(17) \ \ (25)  \\ 
		4.697 (33)(16)    & 11.80   & -3.69(2) \ & 0.002& 0.358(1) & 0.356 (3)(1)  & 4.341  (33)(17) \ \ (50)    \\  
    14.513 \ (6)(16)  & 6.54    & -4.10(3) \ & 0.002& 0.357(1) & 0.356 (3)(1)  & 14.157 \ \ (7)(17) \ \ (24) \\  
		127.461  (13)(16) & 2.20    & -8.71(22) & 0.000& 0.346(1) & 0.346 (3)(1)  & 127.115	  (14)(17) \ \ (31) \\ 
  \end {tabular}
	  \caption{\textbf{Dissociation frequencies and binding energies \cite{Zuerntakenthesis2012}}. The scattering length $a_{12}$ is calculated using equation \ref{universal bound state 1rst order correction} and $a_{13}$ is determined from $a_{13}(B)$ of ref. \cite{Bartenstein2005}. Its systematic error results from the $1\;$G uncertainty of the $\vert13\rangle$ pole in ref \cite{Bartenstein2005}. The confinement shift of the initial and final state of the rf-transition is calculated using equation \ref{implicit relation for E}. The error is the propagated error of $a_{13}$. The difference of both shifts determines the total confinement shift $\nu_{\text{cs}}$. The first parenthesis states the statistical error of the zero point energy of the relative motion given by the SEM of the radial trap frequency which is determined from the separation of the sideband peaks in the dissociation spectra. The second parenthesis gives the systematic error of the confinement shift of the final state. The quadratic addition of both errors determines $\Delta_{\nu_{\text{cs}}}$. The binding energy $E_b$ is calculated from the difference between the measured dissociation frequency and the confinement shift. The first parenthesis gives the statistical error $\sigma_{\nu_{E_b}}$(see TABLE \ref{table free-free and bound-free}). The second parenthesis gives the systemtic error $\Delta_{\nu_{E_b}}=\Delta_{\nu_{\text{model}}}$+$\Delta_{\nu_{\text{density}}}$+$\Delta_{\nu_{\text{cs}}}$. The third parenthesis gives the sum of the statistical and systematic error.}
		  \label{table free-free and bound-free part 2}
	\normalsize
\end {table*} 

\begin{figure} [th!]
\centering
	\includegraphics [width=8.5 cm] {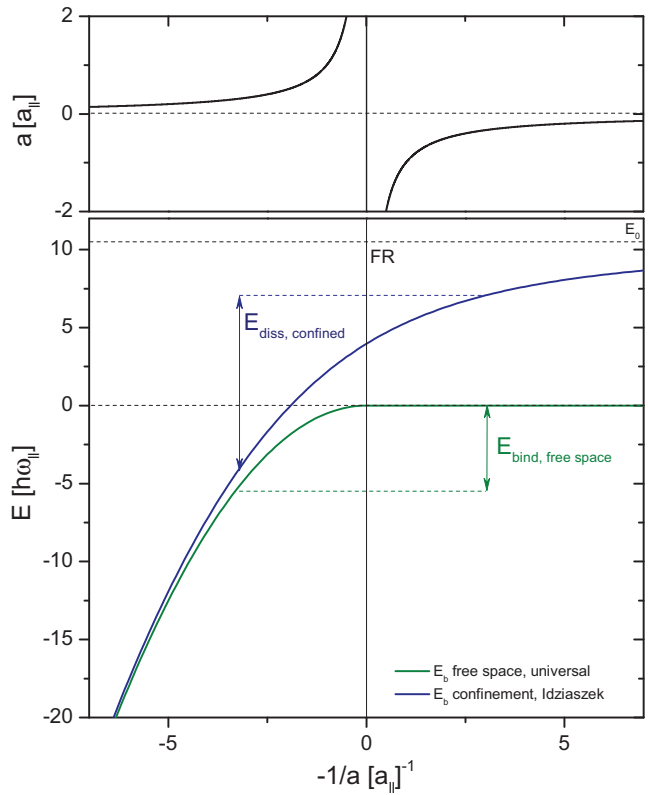}
	\caption{\textbf{Confinement shift \cite{Zuerntakenthesis2012}}. The green curve shows the energy of the universal bound state associated with a Feshbach resonance \cite{Chin2010}. At the point where the scattering length diverges at the Feshbach resonance (FR, upper panel) the universal bound state reaches the continuum. By measuring its binding energy in free space (green arrow) as a function of the magnetic field one can determine the position of the Feshbach resonance. The blue curve shows the universal bound state in the presence of the confinement (\cite{idziaszek2006}, $\eta=10$). The blue arrow indicates an RF transition from a molecule at $a>0$ to two atoms with $a<0$. Due to the confinement the dissociation frequency is shifted with respect to the one in free space. The units are $\omega_\parallel=2\pi\nu_{\text{ax}}$ and $a_\parallel=\sqrt{\frac{\hbar}{\mu \omega_\parallel}}$.}
	\label{fig:comparison-free-confined}
\end{figure}
%---------------------------

\subsection{Confinement shift} %\cite{Zuerntakenthesis2012}}.\\
To determine the binding energy of the molecules we have to subtract the effects of the confining potential from the dissociation frequency $\delta \nu$. The main contribution to the confinement shift is the zero point energy $E_0$ of the relative motion of the dissociated particles in the trap which is given by the frequency $\nu_0=\nu_r+\frac{1}{2}\nu_{ax}=367(3)\,$Hz. In the limit of vanishing scattering length, $a\rightarrow -0\,$, $E_0$ is the only contribution to the shift. For non-zero scattering length an exact expression for the energy of two particles in a cigar shaped harmonic trap with aspect ratio $\eta$ has been derived in ref. \cite{idziaszek2006}. It relates $a$ to the total energy $E<E_0$ of the two particles
%---------------------------
\begin{equation}
-\frac{1}{a}=\frac{1}{\sqrt{\pi}}\,\mathcal{F}\left(-\mathcal{E}/2\right).
\label{implicit relation for E}
\end{equation}
%---------------------------
with $\mathcal{E}=E-E_0$. The integral representation of $\mathcal{F}\left(x \right)$ is given by
%---------------------------
\begin{equation}
\mathcal{F}\left(x \right)=\int^{\infty}_{0} dt \left(\frac{\eta e^{-xt}}{\sqrt{1-e^{-t}}\,(1-e^{-\eta t})} -\frac{1}{t^{3/2}}   \right)
\label{F integral representation} 
\end{equation}
%---------------------------
To calculate the energy we numerically solve equation \ref{implicit relation for E}. The result for an aspect ratio of $\eta=10$ is shown in figure \ref{fig:comparison-free-confined} (blue curve). The difference to the universal bound state in free space (green curve, \cite{Chin2010}) determines the confinement shift. \\
To obtain the confinement shift of the initial state and the final state of the dissociation measurement at different magnetic fields we estimate the scattering length $a_{12}$ from the corresponding dissociation frequencies. Therefore we first estimate the binding energy by subtracting $\nu_0$ from the dissociation frequency. Then we calculate the scattering length using the expression of ref. \cite{gribakin1993} for the energy $E_{\text{gf}}$ of the bound state which considers effective range corrections to first order	
%---------------------------
\begin{equation}
E_{\text{gf}}=\frac{\hbar^2}{\mu (a_{12}-\overline{a})^2}
\label{universal bound state 1rst order correction}
\end{equation}
%---------------------------
with $\mu$ the reduced mass and with the so-called mean scattering length 
%---------------------------
\begin{equation}
\overline{a}\approx 0.487 r_{\text{vdw}}
\label{amean}
\end{equation}
%---------------------------
where $r_{\text{vdw}}$ is the range of the van-der-Waals potential. The scattering length $a_{13}$ at the corresponding magnetic fields is taken from ref. \cite{Bartenstein2005}. The scattering length $a_{12}$  of the initial state and $a_{13}$ of the final state of the rf-dissociation measurement and the corresponding confinement shifts are listed in TABLE \ref{table free-free and bound-free part 2}. By subtracting the total confinement shift from the dissociation frequency we obtain the binding energy of a molecule in free space.

\end{document}